\documentstyle[aps,prl,multicol,epsf]{revtex}
\begin{document}
\newcommand{\tbox}[1]{\mbox{\tiny #1}}
\newcommand{\half}{\mbox{\small $\frac{1}{2}$}}
\newcommand{\mbf}[1]{{\mathbf #1}}


\title{Unification of perturbation theory, RMT and semiclassical
considerations in the study of parametrically-dependent eigenstates}

\author{Doron Cohen and Eric J. Heller}

\date{August 1999, December 1999}

\address{Department of Physics, Harvard University}

\maketitle


\begin{abstract} 
We consider a classically chaotic system that is described
by an Hamiltonian ${\cal H}(Q,P;x)$ where $x$ is a constant
parameter. Our main interest is in the case of a gas-particle
inside a cavity, where $x$ controls a deformation
of the boundary or the position of a `piston'. The
quantum-eigenstates of the system are $|n(x)\rangle$.
We describe how the parametric kernel
$P(n|m)= |\langle n(x)|m(x_0)\rangle|^2$
evolves as a function of $\delta x=x{-}x_0$.
We explore both the perturbative and the non-perturbative regimes,
and discuss the capabilities and the limitations of semiclassical
as well as of random-waves and random-matrix-theory (RMT) considerations.
\end{abstract}

\begin{multicols}{2}

Consider a system that is described by an Hamiltonian
${\cal H}(Q,P;x)$ where $(Q,P)$ are canonical variables and
$x$ is a constant parameter. Our main interest is in the case
where the parameter $x$ represents
the position of a small rigid body (`piston') which is located inside
a cavity, and the $(Q,P)$ variables describe the motion of
a `gas particle'. It is assumed that the system is classically chaotic.
The eigenstates of the quantized Hamiltonian are $|n(x)\rangle$ and
the corresponding eigen-energies are $E_n(x)$.
The eigen-energies are assumed to be ordered, and the
mean level spacing will be denoted by $\Delta$.
We are interested in the parametric kernel
\begin{eqnarray} \label{e_1}
P(n|m) \ = \ |\langle n(x)|m(x_0)\rangle|^2
\ = \ \mbox{trace}(\rho_n\rho_m)
\end{eqnarray}
In the equation above $\rho_m(Q,P)$ and $\rho_n(Q,P)$
are the Wigner functions that correspond to the
eigenstates $|m(x_0)\rangle$ and $|n(x)\rangle$
respectively. The trace stands for $dQdP/(2\pi\hbar)^d$
integration. The difference $x-x_0$ will be denoted by $\delta x$.
We assume a dense spectrum. The kernel $P(n|m)$,
regarded as a function of $n{-}m$, describes an energy
distribution. As $\delta x$ becomes larger, the width as
well as the whole profile of this distribution `evolves'.
Our aim is to study this parametric-evolution (PE).

The understanding of PE is essential for the analysis of
experimental circumstances where the `sudden approximation'
applies \cite{heller}. It also constitutes a preliminary
stage in the studies of quantum dissipation \cite{frc}.
The function $P(n|m)$ has received different names such as
`strength function' \cite{flamb} and `local density of
states' \cite{casati}.
Some generic features of PE can be deduced by referring
to time-independent first-order perturbation theory (FOPT),
and to random-matrix-theory (RMT) considerations \cite{wigner,casati}.
Other features can be deduced using
classical approximation \cite{casati,felix},
or its more controlled version that we are going to
call phase-space semiclassical approximation \cite{frc}.
Still another strategy is to use time-domain
semiclassical considerations \cite{heller}.
In case of cavities one can be tempted to use
`random-wave' considerations as well.
Depending on the chosen strategy, {\em different} results can
be obtained. The `cavity' system is a prototype example
for demonstrating the `clash' between
the various approaches to the problem.


We are considering the cavity example where we have
a `gas' particle whose kinetic energy is
$E=\half mv^2$, where $m$ is its mass, and $v$
is its velocity. The `gas' particle is moving inside
a cavity whose volume is $\mathsf{V}$ and whose
dimensionality is $d$.
The ballistic mean free path is $\ell_{\tbox{bl}}$.
The area of the displaced wall-element
(`piston' for brevity) is $A$, while its
effective area is $A{\tbox{eff}}$,
see \cite{frc} for geometrical definition.
The mean free path $\ell_{\tbox{col}}\approx{\mathsf{V}}/A$
between collisions with the piston may be much
smaller compared with $\ell_{\tbox{bl}}$.
The penetration distance upon a collision is $\ell=E/f$,
where $f$ is the force that is exerted by the wall.
Upon quantization we have an additional length
scale, which is the De-Broglie wavelength
$\lambda_{\tbox{B}}=2\pi\hbar/(mv)$.  We shall distinguish
between the {\em hard walls} case where we assume
$\ell < \lambda_{\tbox{B}} \ll \ell_{\tbox{bl}}$,
and {\em soft walls} for which  $\lambda_{\tbox{B}} \ll \ell$.
Note that taking $\hbar\rightarrow 0$ implies soft walls.


For convenience of the reader we start by listing
the various expressions that can be derived for
$P(n|m)$, along with an overview of our PE picture.
Then we proceed with a detailed presentation.
We are going to argue that four parametric scales
$\delta x_c^{\tbox{qm}}\ll
\delta x_{\tbox{NU}}\ll
\delta x_{\tbox{prt}}\ll
\delta x_{\tbox{SC}}$
are important in the the study of PE.


Standard FOPT assumes that $P(n|m)$ has a simple perturbative structure
that contains mainly one state:
\begin{eqnarray} \label{e_2}
P(n|m) \ \approx \ \delta_{nm} + \mbox{Tail}(n-m)
\end{eqnarray}
We define $\delta x_c^{\tbox{qm}}$ to be the
parametric change that is required in order to mix
neighboring levels.
For $\delta x > \delta x_c^{\tbox{qm}}$ an improved
version of FOPT implies that $P(n|m)$ has a
core-tail structure \cite{frc}:
\begin{eqnarray} \label{e_3}
P(n|m) \ \approx \ \mbox{Core}(n-m) + \mbox{Tail}(n-m)
\end{eqnarray}
The {\em core} consists of those levels that are mixed
non-perturbatively, and the tail evolves as if standard FOPT
is still applicable.
In particular we argue that the tail grows like $\delta x^2$,
and not like $\delta x$. We also explain how the core-width depends
on $\delta x$.  It should be noted that
Wigner's Lorentzian \cite{wigner,casati} can be regarded
as a special case of core-tail structure.

Another strategy is to use {\em semiclassical considerations}.
The simplest idea is to look on the definition
(\ref{e_1}) and to argue that $\rho_n(Q,P)$ and $\rho_m(Q,P)$
can be approximated by microcanonical distributions. This is
equivalent to the classical approximation that has been
tested in \cite{felix}.
If we try to apply this approximation to the cavity example
we should be aware of a certain complication that is illustrated
in Fig.1. One obtains
\begin{eqnarray} \label{e8}
P(n|m)= \left(1{-}\frac{\tau_{\tbox{cl}}}{\tau_{\tbox{col}}}\right)
\delta(n-m)+\mbox{S}\left(\frac{E_n-E_m}{\delta E_{\tbox{cl}}}\right)
\end{eqnarray}
The detailed explanation of this expression is postponed to
a later paragraph. 
A more careful semiclassical procedure is to take the width of
Wigner function into account. Namely,
we can approximate $\rho_n(Q,P)$ and $\rho_m(Q,P)$
by {\em smeared} microcanonical distributions.
It can be used in order to get an idea concerning the
quantum mechanical `interpretation' of the Dirac's delta function
component in (\ref{e8}).  The result is
\begin{eqnarray} \label{e7}
\delta(n-m) \ \ \mapsto \ \ \frac{1}{\pi} \
\frac{\delta E_{\tbox{SC}}}{\delta E_{\tbox{SC}}^2+(E_n{-}E_m)^2}
\end{eqnarray}
with $\delta E_{\tbox{SC}}=\hbar/\tau_{\tbox{bl}}$,
where $\tau_{\tbox{bl}}=\ell_{\tbox{bl}}/v$.
However, we are going to argue that the latter procedure,
which is equivalent to the assumption of having uncorrelated
random waves, is an over-simplification. It is better to use
the {\em time-domain semiclassical approach} which is based on
the realization that $P(n|m)$ is related to the so-called
survival amplitude via a Fourier transform \cite{heller},
leading to the identification
$\delta E_{\tbox{SC}}=\hbar/\tau_{\tbox{col}}$,
where $\tau_{\tbox{col}}=\ell_{\tbox{col}}/v$.

\begin{figure}
\epsfysize=2.35in
\epsffile{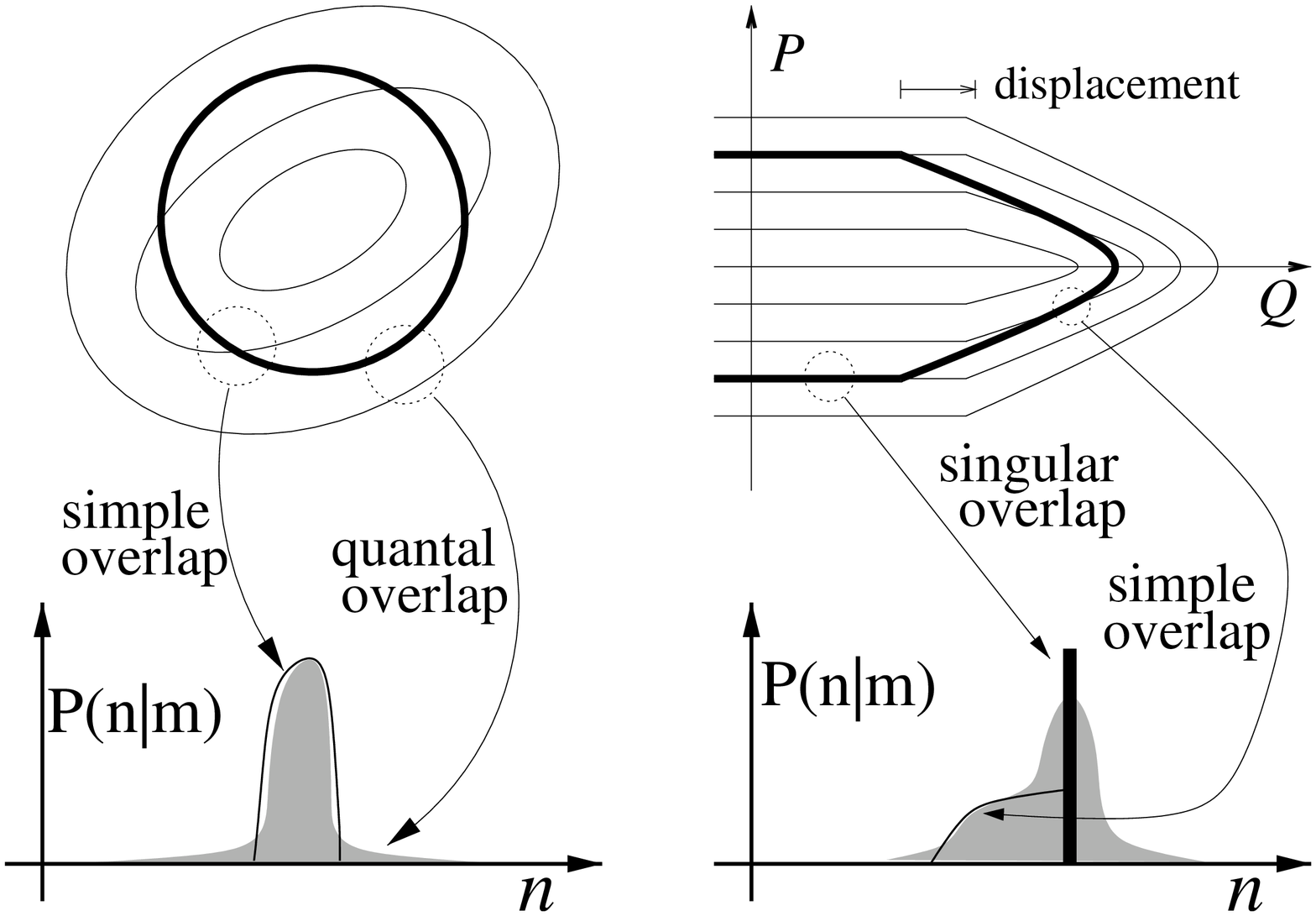}

\noindent
{\footnotesize {\bf FIG. 1.}
Phase space illustration of the energy surface
(represented by bold solid line) that support the
Wigner function of a given eigenstate $|m(x_0)\rangle$,
and the energy surfaces (light solid lines)
that support the Wigner functions of some of
the eigenstates $|n(x)\rangle$.
The left illustration refers to an hypothetical generic
case, while the right illustration refers to the
cavity example. The associated $P(n|m)$ is plotted
below each of the phase space illustrations:
The classical behavior is indicate by the black lines,
and the quantum-mechanical behavior is represented by
the grey filling. It should be realized that
detailed quantal-classical correspondence assumed,
which is guaranteed only if
$\delta x > \delta x_{\tbox{SC}}$.
In the quantum-mechanical case classical sharp-cutoffs
are being smeared (`tunneling' correction).
In the cavity example the classical delta-singularity
is being smeared as well. In the latter case a naive
phase-space picture cannot be used in order to determine
the width of the smearing. }
\end{figure}

The important point to realize is that (\ref{e8})
with (\ref{e7}) is fundamentally different from
either (\ref{e_2}) or (\ref{e_3}). The main purpose
of this Letter is to give a clear idea of the route
from the regime where perturbation theory applies to the
non-perturbative regime regime where semiclassical
consideration become useful.
We are going to explain that the width of the {\em core}
in (\ref{e_2}) defines a `window' through which we can view
the `landscape' of the semiclassical analysis.
As $\delta x$  becomes
larger, this `window' becomes wider, and eventually
some of semiclassical structure is exposed. This is marked
by the non-universal parametric scale $\delta x_{\tbox{NU}}$.
For $\delta x$ much larger than $\delta x_{\tbox{NU}}$,
the non-universal structure (\ref{e7}) of the core is exposed.
Still, the perturbative tail of (\ref{e_3}) may survive
for relatively large values of $\delta x$.
One wonders whether this tail survives
for arbitrarily large $\delta x$. While the answer for
the latter question may be positive for hard walls,
it is definitely negative for soft walls,
as well as for any other generic system.
Assuming soft walls, one should realize that the
perturbative tail of (\ref{e_3}) occupies a finite bandwidth.
It is well known \cite{mario} that having finite bandwidth
is a generic feature of all quantized systems, provided
$\hbar$ is reasonably small.
Therefore one should introduce an additional
parametric scale $\delta x_{\tbox{prt}}$.
For $\delta x\gg\delta x_{\tbox{prt}}$ the {\em core}
spills over the bandwidth of the perturbative {\em tail},
and $P(n|m)$ becomes purely non-perturbative.
The non-perturbative $P(n|m)$ does not necessarily
correspond to the classical approximation (\ref{e8}).
We are going to introduce one more additional scale $\delta x_{\tbox{SC}}$.
For $\delta x\gg\delta x_{\tbox{SC}}$ detailed
quantal-classical correspondence is guaranteed, and
(\ref{e8}) with (\ref{e7}) becomes applicable.


Expression (\ref{e_2}) is a straightforward result of
standard time-independent FOPT where
\begin{eqnarray} \label{e1}
\mbox{Tail}(n-m) \ &=& \
\left|\left(\frac{\partial{\cal H}}{\partial x}\right)_{nm}\right|^2
\frac{\delta x^2}{(E_n{-}E_m)^2}
\end{eqnarray}
An estimate for the matrix elements
$({\partial{\cal H}}/{\partial x})_{nm}$
follows from simple considerations \cite{frc}.
Upon substitution into (\ref{e1}) it leads to:
\begin{eqnarray} \nonumber
P(n|m) \approx 
\left(\frac{\delta x}{\delta x_c^{\tbox{qm}}}\right)^{\beta}
\frac{1}{(n-m)^{2{+}\gamma}}
\\ \label{e4}
\ \ \ \ \mbox{for $b(x)\ll|n-m|\ll b$}
\end{eqnarray} 
with $\beta=2$ and $b(x)=0$. We have defined
\begin{eqnarray} \label{e3}
\delta x_c^{\tbox{qm}} \ \ \approx \ \
\sqrt{\frac{\Gamma((d{+}3)/2)}
{4\pi^{(d{-}1)/2}} \
\frac{1}{A_{\tbox{eff}}} \
\lambda_{\tbox{B}}^{d{+}1} }
\end{eqnarray}
We shall refer to the dependence of
$|({\partial{\cal H}}/{\partial x})_{nm}|^2$ on
$n-m$ as the band-profile. It is well known \cite{mario}
that the band-profile is related (via a Fourier transform)
to a classical correlation function.
If successive collisions with the 'piston' are
uncorrelated then we have $\gamma{=}0$.
But in other typical circumstances \cite{prm}
we may have $0{<}\gamma$.
The matrix $({\partial{\cal H}}/{\partial x})_{nm}$
is not a banded matrix unless we assume soft (rather than hard)
walls. In the latter case the bandwidth
$\Delta_b=(\hbar/\tau_{\tbox{cl}})$ is related to the
collision time $\tau_{\tbox{cl}}=\ell/v$ with the walls.
Having hard walls ($\ell<\lambda_{\tbox{B}}$),
implies that $\Delta_b$ becomes (formally) larger than $E$.
The notion of bandwidth is meaningful only for
soft walls ($\ell\gg\lambda_{\tbox{B}}$).
In dimensionless units the bandwidth it is
commonly denoted by $b=\Delta_b/\Delta$.

The standard result (\ref{e_2}) with (\ref{e1}) of FOPT is valid
as long as $\delta x \ll \delta x_c^{\tbox{qm}}$.
Once $\delta x$ becomes of the order of $\delta x_c^{\tbox{qm}}$,
we expect few levels to be mixed non-perturbatively.
Consequently (for \mbox{$\delta x > \delta x_c^{\tbox{qm}}$})
the standard version of FOPT breaks down.
As $\delta x$ becomes larger, more and more levels
are being mixed non-perturbatively, and it is natural
to distinguish between {\em core} and {\em tail} regions.
The core-width $b(x)$ is conveniently defined as the
participation ratio (PRR),
namely $b(x){=}( \sum_n (P(n|m))^2 )^{-1}$.
The {\em tail} consists of all the levels that become
`occupied' due to first-order transitions from the
{\em core}. It extends within the range $b(x){<}|n{-}m|{<}b$.
Most of the spreading probability is contained within
the core region, which implies a natural extension
of FOPT: The first step is to make a transformation
to a new basis where transitions within the core are
eliminated; The second step is to use FOPT (in the new basis)
in order to analyze the core-to-tail transitions.
Details of this procedure are discussed in \cite{frc},
and the consequences have been tested numerically \cite{prm}.
The most important (and non-trivial) consequence of this
procedure is the observation that mixing on small scales
does not affect the transitions on large-scales.
Therefore we have in the tail region $P(n|m)\propto \delta x^2$
rather than $P(n|m)\propto \delta x$.
The above considerations can be summarized by stating that
(\ref{e4}) holds with $\beta=2$ well beyond the breakdown
of the standard FOPT.

We turn now to discuss the non-perturbative
structure of the core. The identification of
$b(x)$ with the inverse-participation-ratio is
a practical procedure as long as we assume a
simple energy spreading profile where the
core is characterized by a {\em single} width-scale.
As long as this assumption (of having structure-less core)
is true we can make one step further and argue that
\begin{eqnarray} \label{e5}
b(x)\Big|_{\tbox{PRR}} \ = \ 2\pi^2 \left(
\frac{\delta x}{\delta x_c^{\tbox{qm}}}
\right)^{2/(1{+}\gamma)}
\ \ \ \ \mbox{assuming $|\gamma|<1$}
\end{eqnarray} 
The argument goes as follows: Assuming that there is
only one relevant energy scale ($b(x)$) it is implied
by (\ref{e4}) that $P(n|m)$ has the normalization
$(\delta x/\delta x_c^{\tbox{qm}})^2/(b(x))^{1{+}\gamma}$.
This should be of order unity. Hence (\ref{e5}) follows.
The tail should go down fast enough ($\gamma>-1$)
else our `improved' perturbation theory does not hold.
Namely, for $\gamma<-1$ the core-width becomes cutoff
dependent (via its definition as an PRR),
and consequently it is not legitimate to neglect
the `back reaction' for core-to-tail transitions.
The tail should go down slow enough ($\gamma<1$) in order
to guarantee that the core width is tail-determined.
Else, if $\gamma>1$ then the core width is expected to be
determined by transitions between near-neighbor
levels leading to a simple linear behavior
$b(x)=(\delta x / \delta x_c^{\tbox{qm}})$.

Non-perturbative features of $P(n|m)$ are associated
with the structure of the {\em core}.
In order to further analyze the non-perturbative
features of $P(n|m)$ we are going to apply semiclassical
considerations. 
An eigenstate $|n(x)\rangle$ can be represented by
a Wigner function $\rho_n(Q,P)$. In the classical
limit $\rho_n(Q,P)$ is supported by the energy
surface ${\cal H}(Q,P;x)=E_n$. However, unlike
microcanonical distribution, it is further characterized
by a non-trivial transverse structure. One should
distinguish between the `bulk' flat-portions of the
energy-surface (where $Q$ describes free motion),
and the relatively narrow curved-portions
(where $Q$ is within the wall field-of-force).
In the curved-portion of the energy surface
(near the turning points), Wigner function
has a transverse {\em Airy} structure
whose `thickness' is characterized by the energy scale
$\Delta_{\tbox{SC}}=((\hbar/\tau_{\tbox{cl}})^2 E)^{1/3}$.
This latter expression is valid for soft
walls ($\lambda_{\tbox{B}}\ll\ell$). In the hard wall case
($\ell<\lambda_{\tbox{B}}$) it goes to $\Delta_{\tbox{SC}}\sim E$.
Unlike the curved-portions, the `bulk' flat-portions of the
energy surface are characterized by
$\Delta_{\tbox{SC}}=(\hbar/\tau_{\tbox{bl}})$.
Now we consider {\em two} sets of eigenstates,
$|n(x)\rangle$ and $|m(x_0)\rangle$, which are
represented by two sets of Wigner functions
$\rho_n(Q,P)$ and $\rho_m(Q,P)$. The probability
kernel (\ref{e_1}) can be written as
$P(n|m)=\mbox{trace}(\rho_n\rho_m)$.
If $\rho_n(Q,P)$ and $\rho_m(Q,P)$
are approximated by microcanonical distributions,
then $P(n|m)$ is just the projection of the
energy surface that correspond to $m$, on the
``new'' energy surface that correspond to $n$.
This leads to the classical approximation Eq.(\ref{e8}).
In the classical limit $n$ and $m$ become continuous
variables, and Dirac's delta just reflects the
observation that most of the energy surface
(the `bulk' component) is {\em not} affected by changing
the position of the classically-small `piston'.
The second term in (\ref{e8}) has the normalization
$(\tau_{\tbox{cl}} / \tau_{\tbox{col}})$, and corresponds
to the tiny component which is affected by the displacement
of the `piston'. For $\delta x < \ell$ it extends
over an energy range $\delta E_{\tbox{cl}}= f \delta x$,
where $f$ is the force which is exerted on the particle
by the wall. When  $\delta x$ becomes larger than $\ell$
the energy spread becomes of order $E$.

In the quantum-mechanical case we should wonder
whether (\ref{e8}) can be used as an approximation,
and what is the proper `interpretation' of
Dirac's delta function.  It is relatively easy to
specify {\em sufficient} condition for the validity
of the classical approximation. Namely,
the transverse structure of Wigner function
can be ignored if 
$\Delta_{\tbox{SC}} \ll |E_n-E_m| \ll \delta E_{\tbox{cl}}$.
For {\em hard} walls $\Delta_{\tbox{SC}}\sim E$ and
therefore the classical approximation becomes inapplicable.
For {\em soft} walls the necessary condition
$\Delta_{\tbox{SC}} \ll \delta E_{\tbox{cl}}$
is satisfied provided $\delta x$ is large enough.
Namely $\delta x \gg \delta x_{\tbox{SC}}$,
with $\delta x_{\tbox{SC}} = (\ell\lambda_{\tbox{B}}^2)^{1/3}$.

We want to go beyond the classical approximation, and
to understand how the classical Dirac's delta function in (\ref{e8})
manifests itself in the quantum mechanical case.
Thus we are interested in the singular overlap
of the `bulk' components (see Fig.1), and the
relevant $\Delta_{\tbox{SC}}$ for the current discussion
is $\hbar/\tau_{\tbox{bl}}$.
The most naive guess is that the contribution due
to the overlap of `bulk' components becomes non-zero
once $|E_n-E_m| < \Delta_{\tbox{SC}}$.
Equivalently, one may invoke a `random-wave' assumption:
One may have the idea that $|n(x)\rangle$ and $|m(x_0)\rangle$
can be treated as {\em uncorrelated} random-superpositions
of plane-waves. Adopting the random-wave assumption, it is technically
lengthy but still straightforward to derive (\ref{e8})
with $\delta E_{\tbox{SC}}=\hbar/\tau_{\tbox{bl}}$.

The naive phase-space argument that supports the `random wave'
result (\ref{e7}) is definitely wrong.
One should realize that $|E_n-E_m| < \Delta_{\tbox{SC}}$
is a {\em necessary} rather than a sufficient condition
for having a non-vanishing `bulk' contribution.
This latter observation becomes evident if one
considers the trivial case $\delta x=0$ for which
we should get $P(n|m)=0$ for any $n\ne m$.
Thus $\hbar/\tau_{\tbox{bl}}$ should be regarded as an upper
limit for $\delta E_{\tbox{SC}}$.
We are going to argue that the correct result
(for large enough $\delta x$) is indeed (\ref{e7}),
but $\tau_{\tbox{bl}}$ should be replaced by the possibly
much larger length-scale $\tau_{\tbox{col}}$.

In order to go beyond the random-wave assumption we
use the {\em time-domain semiclassical approach} which is based on
the realization that $P(n|m)$ is related to the so-called
survival amplitude via a Fourier transform \cite{heller}:
\begin{eqnarray} \label{e9}
\sum_n P(n|m)2\pi\delta(\omega{-}
\mbox{\small$\frac{E_n}{\hbar}$}) =
{\cal F\! T} \ \langle m |
\exp(-i\mbox{\small $\frac{{\cal H}t}{\hbar}$})
|m\rangle  
\end{eqnarray}
Note that $|m\rangle$ is an eigenstate of
${\cal H}(Q,P;x_0)$ while ${\cal H}={\cal H}(Q,P;x)$.
The knowledge of the short time dynamics, via classical
considerations, can be used to obtain the
`envelope' of $P(n|m)$.
Adopting Wigner's picture, the evolving $|m\rangle$  in the
right hand side of (\ref{e9}) is represented
by an evolving (quasi) distribution $\rho_m(Q,P;t)$.
Let us assume that the `piston' is small, such that the collision rate
with it ($1/\tau_{\tbox{col}}$) is much smaller than $1/\tau_{\tbox{bl}}$.
Due to the chaotic nature of the motion successive collisions
with the piston are uncorrelated. It follows that
the portion of $\rho_m(Q,P;t)$ which is {\em not} affected
by collisions with the `piston' decays exponentially
as $\exp(-t/\tau_{\tbox{col}})$. It is reasonable to assume that
any scattered portion of $\rho_m(Q,P;t)$ lose phase-correlation
with the unscattered portion. Therefore the right
hand side of (\ref{e9}) is the Fourier transform of
an exponential. Consequently $P(n|m)$ should have
the Lorentzian shape (\ref{e7}), but the correct energy-width
is $\delta E_{\tbox{SC}} = \hbar/\tau_{\tbox{col}}$
rather than $\hbar/\tau_{\tbox{bl}}$.

For an extremely small parametric change
such that $\delta x \ll \delta x_c^{\tbox{qm}}$
we have the simple perturbative structure (\ref{e_2}).
Then, for larger values of $\delta x$ the energy distribution
develops a core. As long as this core is structure-less
it is characterized by the single width-scale
$b(x)$ of (\ref{e5}). Now we would like to define
a new parametric scale $\delta x_{\tbox{NU}}$.
By definition, for $\delta x \gg \delta x_{\tbox{NU}}$
non-universal features manifest themselves,
and the core is characterized by more than
one width-scale. For our `cavity' example this
happens once the semiclassical Lorentzian structure (\ref{e7})
is exposed. This happens when $b(x)$ of (\ref{e5})
becomes larger than $\delta E_{\tbox{SC}}/\Delta$,
leading to 
\begin{eqnarray} \label{e10}
\delta x_{\tbox{NU}} \ = \ \frac{1}{4\pi}
\left( (d{+}1)\frac{A}{A_{\tbox{eff}}} \right)^{1/2}
\ \lambda_{\tbox{B}}
\end{eqnarray}
Let us re-emphasize that the semiclassical
argument that is based on (\ref{e9})
applies to the non-universal parametric regime
$\delta x \gg \delta x_{\tbox{NU}}$, where
the semiclassical Lorentzian structure (\ref{e7}) is exposed.
It is also important to realize that
in the non-universal regime we do not have
a theory for the $b(x)$ of (\ref{e4}). The derivation
of (\ref{e5}) is based on the assumption of having
a structure-less core, and therefore pertains only
to the universal regime.

It is well known \cite{mario} that for any quantized system
$({\partial{\cal H}}/{\partial x})_{nm}$  is characterized,
for sufficiently small $\hbar$, by a finite bandwidth $\Delta_b$.
Consequently it is possible to define
a non-perturbative regime $\delta x \gg \delta x_{\tbox{prt}}$,
where the condition $b(x) \ll b$ is violated.
In the non-perturbative regime  expression (\ref{e4})
becomes inapplicable because the core spills over
the (perturbative) tail region.
Thus $P(n|m)$ becomes purely non-perturbative.
Hard walls are non-generic as far as the above
semiclassical considerations are concerned.
In the proper classical limit all
the classical quantities should be held fixed (and finite),
while making $\hbar$ smaller and smaller. Therefore
the proper classical limit implies soft walls
($\lambda_{\tbox{B}}\ll\ell$), leading to finite
bandwidth $\Delta_b=\hbar/\tau_{\tbox{cl}}$.
From (\ref{e5}) it follows that the condition $b(x) \ll b$
is definitely not violated for $\delta x \le \delta x_{\tbox{NU}}$.
Hence we conclude that
$\delta x_{\tbox{prt}}\gg\delta x_{\tbox{NU}}$, but we cannot
give an explicit expression since (\ref{e5}) becomes non-valid
in the non-universal regime.

In the parametric regime
$\delta x_{\tbox{NU}} \ll \delta x \ll \delta x_{\tbox{prt}}$
we have on the one hand $\delta E_{\tbox{cl}} \gg \Delta_b$,
and on the other hand $b(x) \ll b$ by definition.
Therefore we cannot get in this regime a contribution
that corresponds to the second term in (\ref{e8}).
A {\em necessary} condition for the manifestation of
this second term is $\delta x \gg \delta x_{\tbox{prt}}$.
However, it should be realized that having
$\delta x \gg \delta x_{\tbox{prt}}$ is not
a sufficient condition for having
detailed correspondence with the classical approximation.
For our `cavity' example detailed correspondence means
that the whole classical structure of (\ref{e8}) is exposed.
As discussed previously, the {\em sufficient} condition
for having such detailed correspondence is
$\delta x \gg \delta x_{\tbox{SC}}$.
This latter condition is always satisfied
in the limit $\hbar\rightarrow 0$.

We thank Michael Haggerty for his useful comments,
and ITAMP for their support.
       

\end{multicols}
\end{document}